\documentclass[onecolumn]{a-template}
\usepackage{hyperref}
\usepackage{doi}
\usepackage{ragged2e} % for formatting of text into nice shape
\usepackage[version=4]{mhchem} % for writing chemical formula in text (specify version to avoid error message)
\usepackage[english]{babel}
\usepackage{lipsum}
\begin{document}

\pagestyle{fancy}
\title{Strain-Driven Thermal and Optical Instability in Silver/Amorphous-Silicon Hyperbolic Metamaterials}
\maketitle
\author{Jose L. Ocana-Pujol$^{*1}$,}
\author{Lea Forster$^{1,2}$,}
\author{Ralph Spolenak$^1$,}
\author{Henning Galinski$^{*1}$.}
\begin{affiliations}
\\~\\
$^1$Laboratory for Nanometallurgy, Department of Materials, ETH Zurich, Z\"urich, Switzerland\\
$^2$Laboratory of Multifunctional Ferroic Materials, Department of Materials, ETH Zurich, Z\"urich, Switzerland\\
$^*$corresponding authors, E-mail: 
{\tt joseo@ethz.ch} and {\tt henningg@ethz.ch}
\end{affiliations}
\begin{abstract}
\begin{justify}Hyperbolic metamaterials show exceptional optical properties, such as near-perfect broadband absorption, due to their geometrically-engineered optical anisotropy. Many of their proposed applications in thermophotovoltaics or radiative cooling, require high-temperature stability. In this work we examine \ce{Ag}/\ce{a-Si} multilayers as a model system for the thermal stability of hyperbolic metamaterials.
Using a combination of nanotomography, finite element simulations and optical spectroscopy, we map the thermal and optical instability of the metamaterials. Although the thermal instability initiates at 300~$^\circ$C, the hyperbolic dispersion persists up to 500~$^\circ$C. Direct finite element simulations on tomographical data provide a route to decouple and evaluate interfacial and elastic strain energy contributions to the instability. Depending on stacking order the instability's driving force is either dominated by changes in anisotropic elastic strain energy due thermal expansion mismatch or by minimization of interfacial energy. Our findings open new avenues to understand multilayer instability and pave the way to design hyperbolic metamaterials able to withstand high temperatures.\end{justify}
\end{abstract}
\begin{justify}
\section{Introduction}
Hyperbolic metamaterials (HMMs) are periodic designer materials with controllable uniaxial optical anisotropy, where the two components of the permittivity tensor can be engineered to have opposite signs, i.e.\ $\epsilon_{\perp}\cdot\epsilon_{\parallel}<0$. This extreme anisotropy offers an unusual degree of freedom to direct the energy flow of light and gives rise to exotic phenomena, such as abnormal refraction~\cite{hoffman_negative_2007,maas_negative_2014}, super-collimation ~\cite{sreekanth_super-collimation_2019,caligiuri_dielectric_2016}, imaging beyond the diffraction limit~\cite{liu_far-field_2007,lu_hyperlenses_2012}, and the realisation of near-perfect broadband absorbers~\cite{narimanov_reduced_2013, wang_perfect_2018}.
\\~\\
The ascribed control of the optical anisotropy enables to realize different dispersion regimes which for a given frequency manifest as optical topological transitions (OTTs) in the isofrequency surface~\cite{krishnamoorthy_topological_2012}. The four optical phases or dispersion regimes given by the isofrequency surface topologies are: an effective dielectric ($\epsilon_{\perp} > 0$ , $\epsilon_{\parallel} > 0$), a Type I HMM ($\epsilon_{\perp} < 0$ , $\epsilon_{\parallel} > 0$), a Type II HMM ($\epsilon_{\perp} > 0$ , $\epsilon_{\parallel} < 0$), and an effective metal ($\epsilon_{\perp} < 0$ , $\epsilon_{\parallel} < 0$). They are commonly described within the effective medium approximation~(EMA)~\cite{cortes_quantum_2014}. For example, in the case of a transition from a dielectric to a Type II HMM the isofrequency surface transforms from a spheroid to a hyperbole. In the absence of loss, the volume of the hyperbolic shell is infinite which feeds back into an infinite local density of optical states~(LDOS)~\cite{poddubny_hyperbolic_2013}.
\\~\\
Multilayers are the most common realization of HMMs~\cite{chang_design_2021}. Of special interest is the interaction of temperature with multilayer HMMs, either in wavelength-switchable  temperature-actuated hyperlenses~\cite{zhang_tunable_2020} and super absorbers~\cite{kbehera_reconfigurable_2021} or in applications with high operating temperatures such as thermophotovoltaics~\cite{dyachenko_controlling_2016,kim_thermal_2019},  hyperthermia therapy~\cite{maccaferri_hyperbolic_2019,zhao_hyperbolic_2021}, or radiative cooling~\cite{raman_passive_2014,chae_spectrally_2020,liu_non-tapered_2019}. However, one of the key challenges for all of these proposed applications is thermal stability. Therefore, it is crucial to understand how temperature changes the geometry and related optical properties of HMMs.
\\~\\
Here, we examine magnetron sputtered silver/amorphous-silicon (\ce{a-Si}) multilayers as a model system to study the thermal stability of HMMs. Silicon deposited by magnetron sputtering, which is suitable for large-scale photonics, is typically amorphous. Amorphous silicon is a different allotrope with different properties than the more widely studied crystalline silicon~(\ce{c-Si}). The \ce{Ag}/\ce{Si} system is characterized by the absence of a low temperature eutectic~\cite{olesinski_ag-si_1989} and very low boundary solubility on the two sides of the phase diagram~\cite{rollert_solubility_1987,weber_equilibrium_2002}. To the best of our knowledge, a complete \ce{Ag}/\ce{a-Si} phase diagram does not exist, but silver diffusivity is known to be slower than in crystalline silicon~\cite{coffa_crucial_1992}. Amorphous silicon also shows distinct optical properties~\cite{thutupalli_optical_1977,pierce_electronic_1972}, which has been exploited in the field of solar energy~\cite{kang_crystalline_2021,sai_impact_2018}. Silver is considered the preferred metal for HMMs applications in the visible due to its low losses~\cite{cortes_quantum_2014}. Moreover, magnetron sputtered stacked \ce{Ag}/\ce{Si} layers were one of the first systems for which hyperbolic behavior was demonstrated~\cite{lu_enhancing_2014}.
\\~\\
In this work, we provide a detailed analysis of the thermal stability of layered \ce{Ag}/\ce{a-Si} HMMs. First, the design and optical properties of the structure is presented. Then, we examine the structural and optical changes of the system upon annealing. The driving forces behind the multilayer instability are examined. Finally, we explore design parameters to test the scope of our findings.
\section{Results and Discussion}
\subsection{Optical Design and Characterization}
A schematic of the hyperbolic metamaterials realized in this work is illustrated in \textbf{Figure~\ref{fig:1}}a. HMMs were synthesized by magnetron sputtering keeping the bilayer unit cell size $\Lambda$~=~40~nm constant for all data shown in Figure~\ref{fig:1} to \ref{fig:3}. The samples were sputtered on a \ce{c-Si} wafer with 50~nm \ce{SiO2} and 50~nm \ce{Si_{x}N_y} as thermal barriers. The film thickness in the multilayer was confirmed by milling and imaging cross-sections using focused ion beam assisted scanning electron microscopy (FIB-SEM). Figure~\ref{fig:1}b reports the optical phase diagram in the visible wavelength range. The values of $\epsilon_{\perp}$ and $\epsilon_{\parallel}$ for selected geometries are given in the Supporting Information (Figure~S2g,b). To include non-local effects~\cite{ishii_non-local_2015}, the optical phase diagram was obtained by FEM simulations and parameter retrieval along the directions parallel and perpendicular to the surface~\cite{smith_electromagnetic_2005}.
\begin{figure*}[t]
  \includegraphics[width=\linewidth]{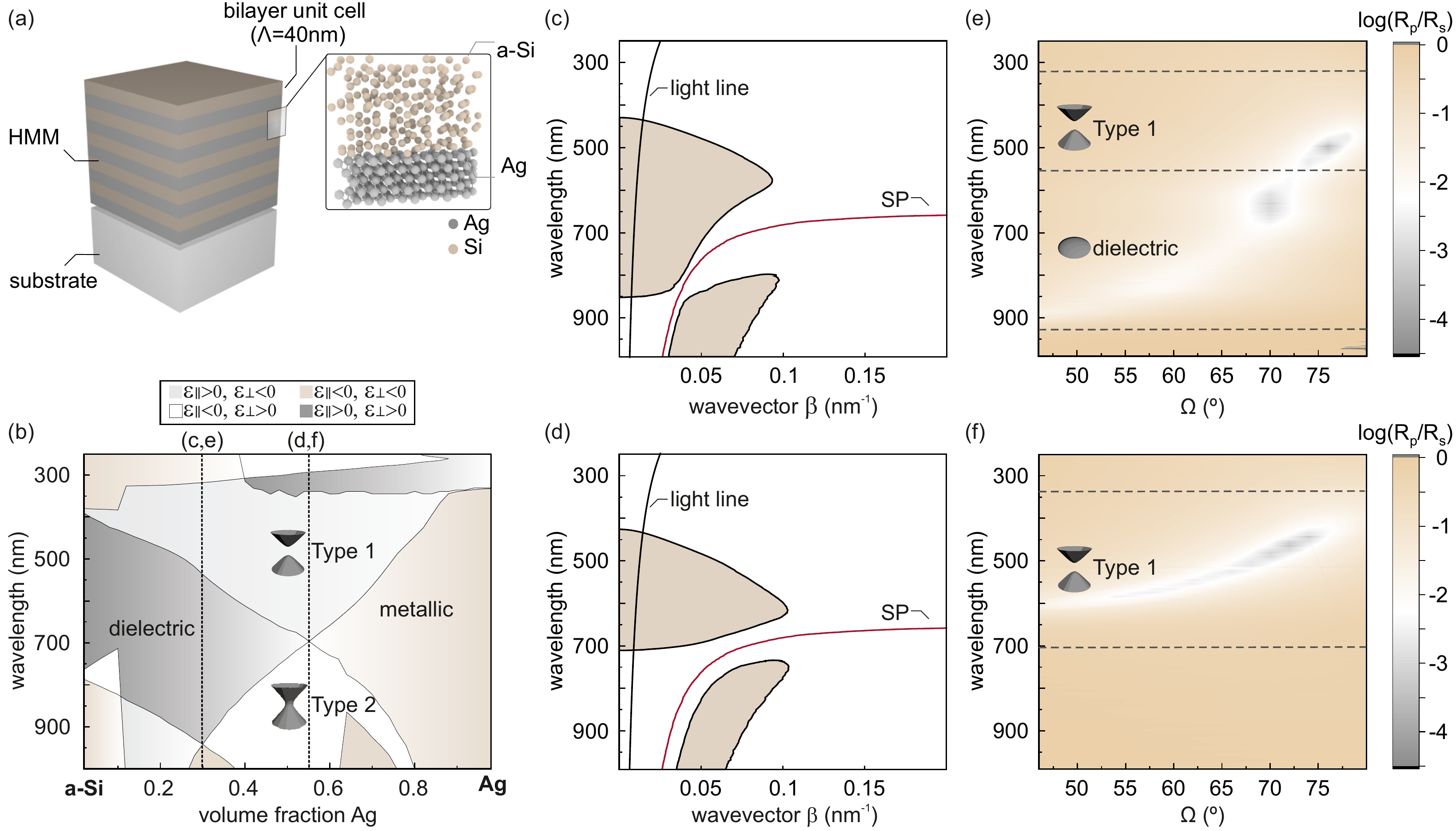}
  \caption{\textbf{Design of the hyperbolic metamaterial} (a) Schematic of a deposited Ag/a-Si multilayer. (b) Optical phase diagram for \ce{Ag}/\ce{a-Si} HMMs with $\Lambda$~=~40~nm unit cell. The color code depicts the different zones: effective dielectric ($\epsilon_{\perp}>0$,$\epsilon_{\parallel}>0$), effective metal ($\epsilon_{\perp}<0$,$\epsilon_{\parallel}<0$), Type I HMM ($\epsilon_{\perp}<0$,$\epsilon_{\parallel}>0$), and Type II HMM ($\epsilon_{\perp}>0$,$\epsilon_{\parallel}<0$). The dashed vertical lines illustrate the expected compositions of the samples with 30~vol.\% \ce{Ag} (for (c) and (e)) and 55~vol.\% \ce{Ag} (for (d) and (f)). (c) and (d) Dispersion relationship of the bulk plasmonic modes of the (c) 30~vol.\% \ce{Ag} and (d) 55~vol.\%\ce{Ag} systems. The black continuous line indicates the light line in vacuum and the red line shows the dispersion relation of surface plasmons at a single \ce{Ag}/\ce{a-Si} interface. (e) and (f) Measured ratio between the p and s-polarized reflectance of the (e) 30~vol.\% \ce{Ag} and (f) 55~vol.\% \ce{Ag} systems.}
  \label{fig:1}
\end{figure*}\\~\\
The calculated plasmonic dispersion of HMMs with 30 vol.\% and 55 vol.\% \ce{Ag} are shown on Figure~\ref{fig:1}c,d, respectively. The graphs show radiative modes left of the surface plasmon dispersion (SP) which intersect with the light line and modes at higher wavevector~$\beta$ and wavelength~$\lambda$ which do not intersect with the light line and can be considered non-radiative~\cite{schilling_uniaxial_2006}. Figure~\ref{fig:1}c,d further indicates that radiative modes exist in both the dielectric and Type I regimes.\\~\\ 
We observe the transition between these optical topologies experimentally by measuring the reflectance ratio $R_{p}/R_{s}$ as a function of incidence angle $\Omega$, since the propagation of hyperbolic modes is limited to p-polarized light. The colormaps in Figure~\ref{fig:1}e,f show minima in the reflectance ratio for Type I and dielectric optical topologies in accordance with Figure~\ref{fig:1}b. Due to the non-radiative nature of the plasmonic modes shown in Figure~\ref{fig:1}c and d, no reflection anisotropy occurs at the Type II HMM and metallic regimes. Moreover, in the case of a system with 30~vol.\%~\ce{Ag} (see Figure~\ref{fig:1}e), the transition between Type I HMM and the dielectric phases should lead to a discontinuity in the pseudo-Brewster angle~\cite{hoffman_negative_2007,cho_experimental_2021}. In agreement with the optical phase diagram (Figure~\ref{fig:1}b), our measurements reproduce this discontinuity shown in Figure~\ref{fig:1}e caused by the change of sign in $\epsilon_{\perp}$ at $\lambda$~=~550~nm. 
\subsection{Thermal and Optical Instability}
\begin{figure*}[t]
  \includegraphics[width=\linewidth]{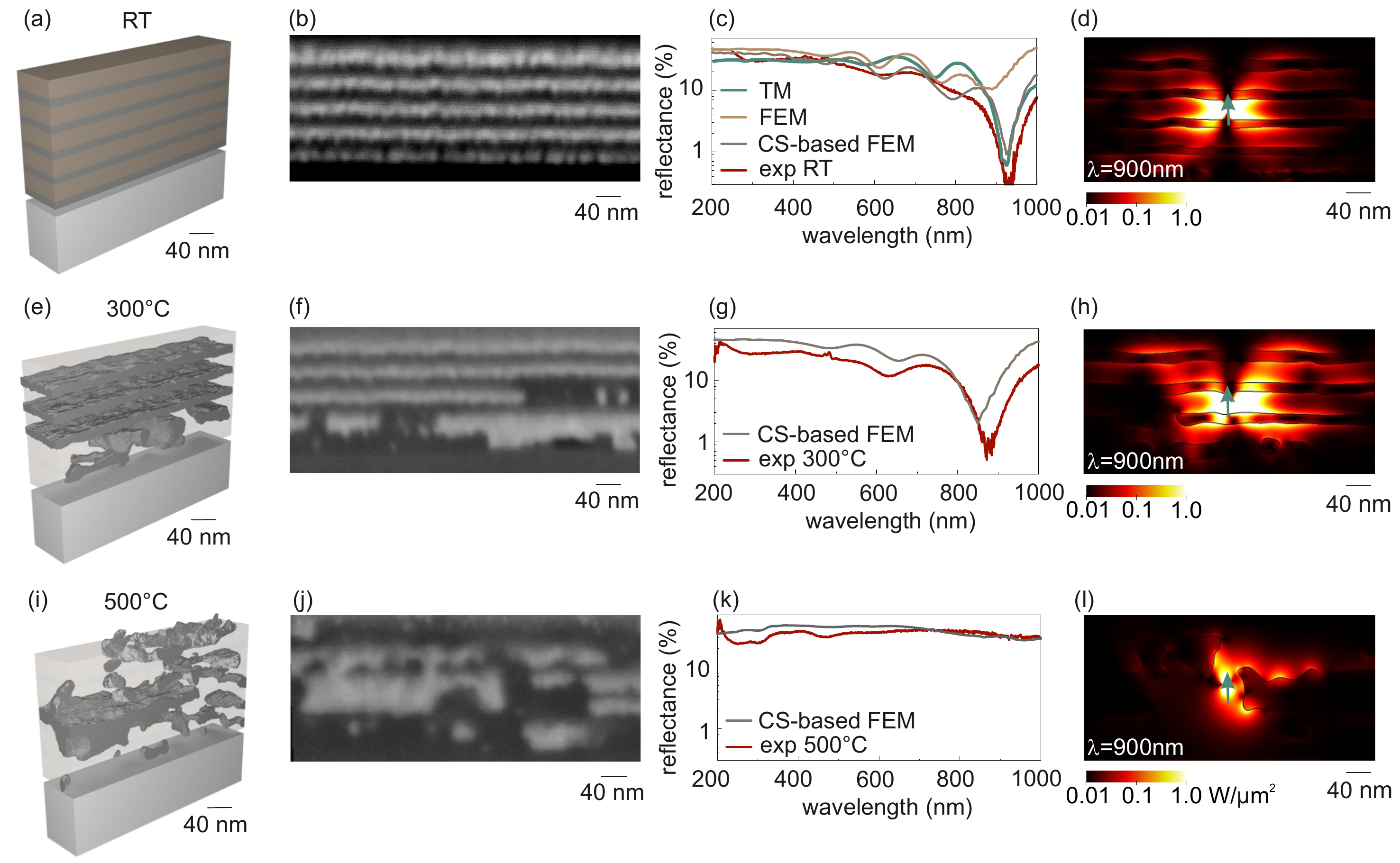}
  \caption{\textbf{Thermal degradation} (a): Three-dimensional (3D) model of the as-deposited structure with 30~vol\% \ce{Ag}. Silver is depicted in gray. (e) and (i): 3D reconstructed FIB nanotomography of the samples annealed at (e) 300~$^\circ$C and (i) 500~$^\circ$C. The gray structure corresponds to the silver phase. (b), (f), and (j) FIB-SEM cross-sections of the as-deposited (b), annealed at 300~$^\circ$C (f), and 500~$^\circ$C state (j). The bright color corresponds to silver.  (c), (g), and (k) Experimental and simulated reflectance at normal incidence in the as-deposited state (g), after annealing at 300~$^\circ$C (h), and 500~$^\circ$C (k). FEM stands for finite element simulations assuming perfectly flat layers, CS-based FEM stands for FEM simulations when the geometry is imported from FIB-SEM cross-sections, and TM is the transfer matrix calculated reflectance including interfacial roughness.  The shown simulated reflectance spectra in (g) and (k) represent an average reflectance obtained by averaging over 8 different cross-sections of the HMM.  (d), (h) and (l) FEM simulation of the electromagnetic energy flow created by a point dipole (with its position and orientation depicted by an arrow) placed inside HMMs with different degree of disorder. The geometries are based on FIB-SEM cross-sections.}
  \label{fig:2}
\end{figure*}
\textbf{Figure~\ref{fig:2}} shows the optical and structural response of the 30~vol.\%~\ce{Ag} multilayers as a function of temperature. The samples were annealed in vacuum at 300~$^\circ$C and 500~$^\circ$C for 1~h. A schematic of the as-deposited structure is shown in Figure~\ref{fig:2}a, while Figure~\ref{fig:2}b illustrates a focused ion beam~(FIB) polished cross-section~(CS) of the as-deposited metamaterial. Figure~\ref{fig:2}c reports a comparison of the reflectance spectrum measured at normal incidence of the as-deposited HMM with several calculated reflectance spectra. Interestingly, the FEM simulations using a geometry without interfacial roughness (see also Figure~\ref{fig:2}a) do not reproduce the measured reflectance dip at 900~nm, while the reflectance spectrum obtained on the basis of the FIB cross-section does. 
\\~\\ 
To understand the origin of this Lorentz-like dip in reflectance of the as-deposited state, we employed the transfer-matrix (TM) method to model the reflectance of the HMM. Scalar scattering theory~\cite{szczyrbowski_determination_nodate} was used to account for contributions of interfacial roughness (see Supporting Information). Figure~\ref{fig:2}c shows the calculated response accounting for roughness reproduces the experimental spectral response. This indicates that the reflectance dip at about 900~nm is caused by coupling to the propagating modes with high-k wavevector. In order to test whether this modes are hyperbolic, we placed a vertical oriented dipole inside the 3D tomography and simulated the energy flow, as shown in Figure~\ref{fig:2}d. The characteristic cone-like pattern of the propagation is a fingerprint of hyperbolic behavior~\cite{potemkin_green_2012}. Furthermore, the energy flow is confined to the multilayer indicating the non-radiative nature of the mode. This result is in line with a previous theoretical work, which identified disorder, such as roughness, in an HMM as viable route to  effectively couple to non-radiative highly absorbing modes~\cite{andryieuski_rough_2014}.
\\~\\ 
When subjected to thermal annealing in ultra-high vacuum, our HMMs exhibit two regimes. Upon annealing at 300~$^\circ$C, the geometry of the HMM is no longer preserved. In both, the FIB nanotomography (Figure~\ref{fig:2}e) and the corresponding cross-sectional image (Figure~\ref{fig:2}f), a redistribution of the lower three Ag-layers is evident. This reorganization is characterized by silver diffusion away from the HMM/substrate interface and by agglomeration and grain growth within the silver phase. Interestingly, the loss in structural integrity is not paralleled by a significant degradation of the optical properties. Despite the structural disorder in the HMM, the metamaterial is robust and exhibits still the characteristic dip in the reflectance spectrum (Figure~\ref{fig:2}g) and cone-like electromagnetic energy flow (Figure~\ref{fig:2}h).
\\~\\ 
Upon annealing at 500~$^\circ$C, the geometry and distinct optical response of the HMM is lost. The FIB nanotomography (Figure~\ref{fig:2}i) and the corresponding cross-sectional image (Figure~\ref{fig:2}j), show a dissolution of the multilayer and the formation of a complex Ag/a-Si composite in its place. The composite contains an interconnected silver phase mostly concentrated in center of the former HMM. Locally, we observe Ag segregation to the surface. At this degree of disorder, the optical response of the HMM is no longer present and the measured and simulated reflectance spectra exhibit a quasi-flat response (Figure~\ref{fig:2}k). In addition, the electromagnetic flow in the geometry shown in Figure~\ref{fig:2}l does no longer feature the characteristic cone-like shape, confirming that the system has lost its hyperbolic dispersion at this temperature.
\\~\\ 
Generally, the inhomogeneous redistribution of silver in the HMM can be understood in the context of a structural instability triggered by temperature~\cite{sridhar_multilayer_1997,lewis_stability_2003}. Here, we found the onset of this instability to be independent of the unit cell's composition and to occur approximately at 300~$^\circ$C for all experimentally analyzed HMM-designs (Figure~S5 in the Supporting Information). Our experiments further show that the HMM/substrate interface has a destabilizing role enhancing the kinetics of the fundamental processes in vicinity of this interface. This is an important result as commonly the reorganization of matter due to a thermal or multilayer instability is assumed to be isotropic~\cite{BELLON2021117041}.
\subsection{Thermodynamics of the Instability}
Motivated by the experimentally observed anisotropic instability, we aim to identify the main contributions to its driving force. Intuitively, we can link the tendency of reorganization or pattern formation in an inhomogeneous mixture, here the Ag/a-Si multilayer, to the lowering of the systems total free energy~\cite{StenderEneGalinskiSchmitz+2008+480+486}. With reference to our layered HMM, we can formulate its total free energy $F$ within the Cahn–Hilliard approach~\cite{CAHN1962179,garcke_cahn-hilliard_2001},
\begin{equation}
    F(c,\vec{u})=\int_{V}f(c) + \kappa (\nabla c) ^2 + E_\text{el}(c,\vec{u})dV.
    \label{eq:1}
\end{equation}
Here, the thermodynamic potential has three main contributions, $f(c)$ is the free energy density of the homogeneous system, $\kappa(\nabla c) ^2$ considers contributions from gradients in composition, e.g.\ interfaces, while $E_\text{el}(c,\vec{u})$ is the anisotropic elastic strain energy density. $c$ is the local concentration and $\vec{u}$ is the displacement vector. We assume that $ E_\text{el}(c,\vec{u})$ originates at the HMM/substrate interface and results from the mismatch in coefficient of thermal expansion (CTE).
\\~\\ 
Taking advantage of our FIB nanotomographies, we performed a series of 3D structural mechanics FEM simulations, as illustrated in \textbf{Figure~\ref{fig:3}}, to determine main contributions of $\kappa(\nabla c) ^2$ and $E_\text{el}(c,\vec{u})$ to the free energy change as function of temperature. Since silver and silicon have a positive enthalpy of formation~\cite{olesinski_ag-si_1989}, we can exclude chemical reactions in the system and assume that $f(c)$ does not contribute to the free-energy change~\cite{lewis_stability_2003}.
\begin{figure*}[t]
  \includegraphics[width=\linewidth]{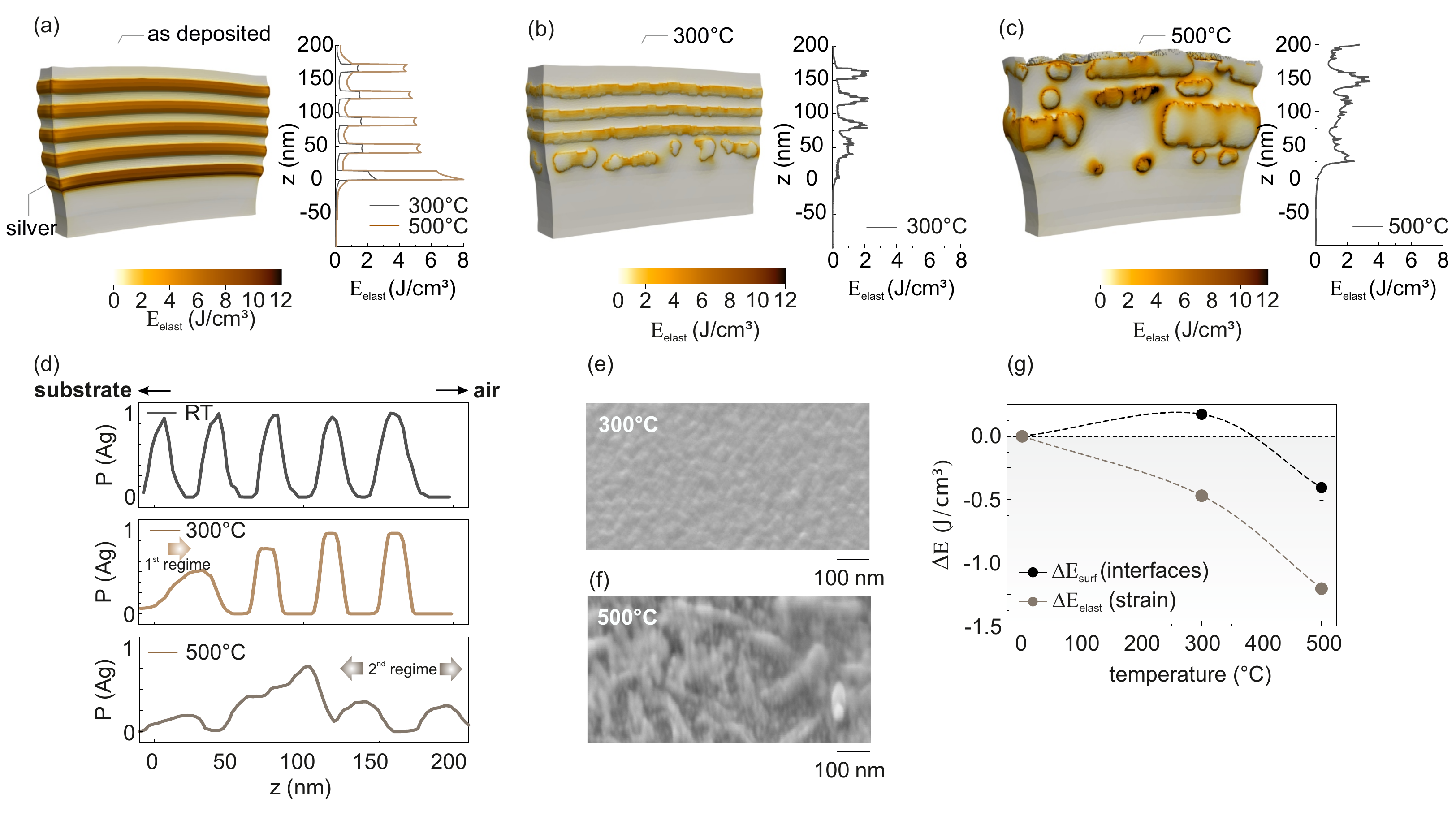}
  \caption{\textbf{Analysis of the driving forces of the instability} (a), (b), and (c) FEM simulations of the elastic strain energy density of the geometries at different temperatures, where the deformation of the geometries maps the thermal expansion. In (a) the nominal HMM design was simulated at both 300~$^\circ$C and 500~$^\circ$C. In panels (b) and (c) the elastic strain energy density are modelled by using the experimental FIB-nanotomography data of the annealed state.  The graphs on the right show the mean values of the elastic strain energy density probed in different locations. (d) \ce{Ag} probability profiles derived from experimental FIB-CS. (e) and (f) SEM images highlighting the change in surface topography (300~$^\circ$C and 500~$^\circ$C) due to silver segregation at 500~$^\circ$C. (g) Average specific energy change $\Delta E$ of the elastic strain energy density and interfacial energy density as function of annealing temperature. The change in elastic strain energy density is negative within the given temperature range, confirming its status as pre-dominant driving force of the instability.}
  \label{fig:3}
\end{figure*}
\\~\\
Figure~\ref{fig:3}a,b,c show the elastic strain energy at different temperatures. In Figure~\ref{fig:3}a, the elastic strain energy caused by the thermal expansion mismatch of Ag and a-Si was simulated assuming nominal flat layers both at 300~$^\circ$C and 500~$^\circ$C with the latter being shown in the color scheme. The subset on the right shows that the elastic strain energy in the metal layer is approximately five times larger than in the dielectric, while the elastic strain energy in the metal decays exponentially from the HMM/substrate interface.
\\~\\
The thermal expansion and elastic strain energy density of the thermally degraded multilayers, shown in Figure~\ref{fig:3}b and \ref{fig:3}c are calculated based on the experimental FIB nanotomographies of the annealed state. Overall, the redistribution of the silver phase significantly lowers the elastic strain energy density. A comparison between the subsets in Figure~\ref{fig:3}a,b,c shows a lower and more homogeneously distributed elastic strain energy within the degraded HMMs. Figure~\ref{fig:3}c illustrates the concentration of \ce{Ag} in the center of the HMM when annealed at 500~$^\circ$C, which leads to a significant reduction of the elastic strain energy. Partial segregation of \ce{Ag} to the HMM/ambient interface enhances this effect (see Figure~\ref{fig:3}e and f).  
\\~\\
To illustrate the reorganization of the silver phase, Figure~\ref{fig:3}d reports the averaged silver depth profiles as function of temperature derived from classified FIB cross-sectional images. Interestingly, the silver distribution and profiles of the elastic strain energy show a similar thermal response. While in the first regime at 300~$^\circ$C redistribution is limited to the HMM/substrate interface, the second regime includes agglomeration in the center of the HMM and partial segregation. This dynamics can be explained by considering the diffusion length of \ce{Ag} in a-Si (Figure~S8) which exceeds the thickness of the HMM for temperatures larger than 425~$^\circ$C.
\\~\\
To determine the main driving force of the instability, we calculated the difference in elastic strain energy $\Delta E_\text{elast}$ and interfacial energy $\Delta E_\text{surf}$. The change in elastic strain energy is derived directly from the FEM simulations, while the interfacial energy is given by the product of the interfacial area derived from the 3D FIB-nanotomographies times the Ag/a-Si interfacial energy (Figure~S8 in the Supporting Information). Figure~\ref{fig:3}g shows the change of this two main contributions to the free energy as function of temperature. The net change in free energy is negative for the analyzed temperatures, i.e.\ the observed instability in fact reduces the system free energy.\\ Notably, at 300~$^\circ$C, the change in elastic strain energy is negative, while the change in interfacial energy is positive, suggesting that the lowering in elastic strain energy is the main driving force of the observed instability. At 500~$^\circ$C both contributions are negative, although the change in elastic strain energy being still pre-dominant.
\\~\\
Our analysis underlines the importance of anisotropic contributions to the thermodynamic potential that describes thermal instabilities. Here, the anisotropic elastic strain energy due to thermal expansion mismatch is identified as the predominant driving force. Most previous literature on multilayer instabilities disregards this effect and treats the morphological redistribution in the context of interfacial energy minimization, including work on sputtered \ce{Ag}/\ce{a-Si} multilayers~\cite{zhao_investigation_1999,KaptaDegradation},  other systems with high CTE missmatch~\cite{misra_effects_2005,cunningham_unraveling_nodate,noauthor_crc_2016}, predictive models~\cite{wan_predictive_2012,lewis_stability_2003} and reviews~\cite{saenz-trevizo_nanomaterials_2020}. It should be noted that these conclusions are often based on experiments using transmission electron microscopy (TEM). Due to sample preparation these kind of experiments occur in the absence of a substrate and thus most likely fall short to reproduce the dynamics observed in real world materials.
\subsection{Influence of Design and Stacking Order}
\begin{figure*}[t]
  \includegraphics[width=\linewidth]{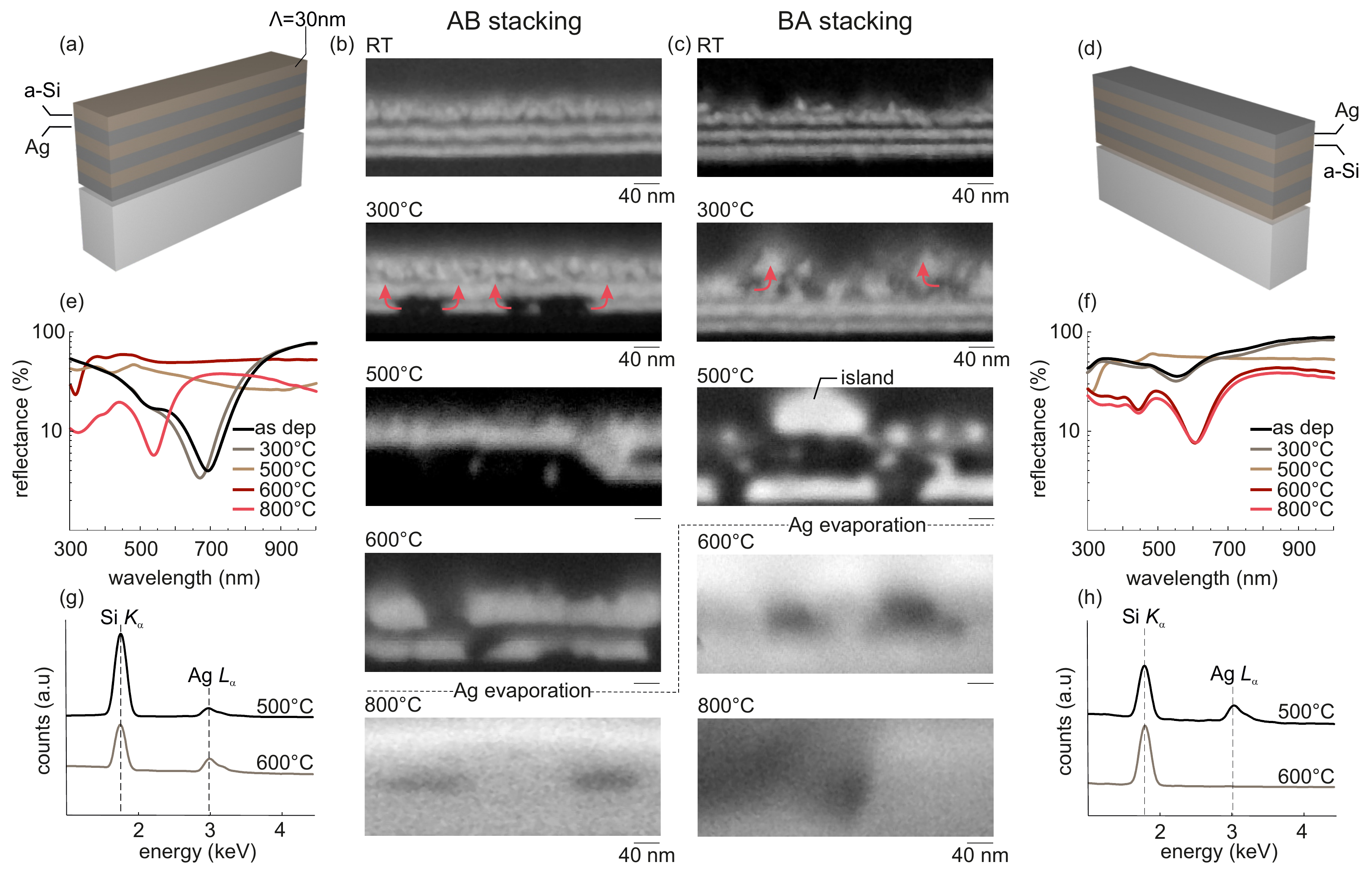}
  \caption{\textbf{Influence of stacking order} (a) and (d): Schematic showing the difference in stacking order between the ABABAB (\ce{Si} on top) and the BABABA (\ce{Ag} on top) architectures. (b) and (c): FIB-SEM cross-sections of the 50~vol.\% \ce{Ag} multilayer with a bilayer unit cell $\Lambda$~=~30~nm annealed at different temperatures.  The scale bar is 40~nm. (e) and (f): Reflectance at normal incidence of the samples annealed at different temperatures. (g) and (h): EDX results showing the presence of the \ce{Ag} and \ce{Si} peaks.}
  \label{fig:4}
\end{figure*}
Based on our previous analysis, this section focuses on the question,whether the driving forces can be controlled by changing the multilayer design parameters. Hence, we changed the HMM design and investigated two samples with six unit cells ($\Lambda$~=~30~nm) and 50~vol.\% \ce{Ag} but different stacking order.
\\~\\
\textbf{Figure~\ref{fig:4}}a,d illustrate the design principle. The AB architecture shown in Figure~\ref{fig:4}a exhibits the stacking order of the samples previously examined in Figure~\ref{fig:1},\ref{fig:2} and \ref{fig:3}, where the layer in contact with the substrate is silver. Figure~\ref{fig:4}b shows that, despite the design change, the structural evolution of the sample with the AB stacking order is in agreement with the instability observed in Figure~\ref{fig:2}: the onset of the instability at 300~$^\circ$C occurs at the bottom interface and the multilayer has disappeared at 500~$^\circ$C.
\\~\\
In contrast, in case of an BA architecture (Figure~\ref{fig:4}d) where amorphous silicon is the material in contact with the substrate, the instability occurs in the vicinity of the HMM/ambient interface (Figure~\ref{fig:4}c). We attribute this change to lower elastic strain energy accumulation in the lower interface due to the similarity in the CTE of \ce{a-Si} and the substrate. In this context, the observed degradation on the top of the multilayer is likely dominated by interfacial energy minimization, which leads to island formation as known from phenomena such as dewetting~\cite{galinski_time-dependent_2013}.
\\~\\
The changes in the optical properties caused by degradation are shown by means of reflectance spectra at normal incidence in Figure~\ref{fig:4}e,f for the AB and BA architectures, respectively. Upon annealing, we observe the persistence of the HMM behavior at 300~$^\circ$C, in line with the results shown in Figure~\ref{fig:2}. This indicates the robustness of optical phases of the HMM independent of the explored design parameters. At 500~$^\circ$C the optical properties change and both systems show a flat reflectance. We attribute this effect to loss of hyperbolic behavior due to multilayer degradation, as observed in Figure~\ref{fig:2}l. It is to note, that we observe hyperbolic behavior in a system composed of only 3 unit cells, which is below the minimum number of components proposed to achieve hyperbolic behavior~\cite{cortes_quantum_2014,ferrari_hyperbolic_2015,ishii_sub-wavelength_2013,li_few-layer_2021}. For further insight into the optical response of these samples, the reader is referred to Figure~S6 in the Supporting Information.
\\~\\
 To test the limits of the second regime, we annealed the samples for 12~h in a vacuum at 600~$^\circ$C and 800~$^\circ$C. Figure~\ref{fig:4}b,c show that, after annealing at these temperatures, the contrast of the SEM images has dramatically changed.  We attribute it to \ce{Ag} evaporation, which is confirmed by the EDX spectra on Figure~\ref{fig:4}g,h. This phenomenon leads to lower reflectivity, as shown in Figure~\ref{fig:4}e,f. The observed evaporation can be understood as a third regime of the thermal instability.
\\~\\
Interestingly, our experiments further show that one can impact the evaporation temperature by the stacking order. While in  samples with BA stacking order silver evaporates at 600~$^\circ$C or 0.7 $T/T_m$ (Figure~\ref{fig:4}h), the AB geometry still contains silver after the same treatment (Figure~\ref{fig:4}g) and evaporation is not observed until 0.85 $T/T_m$~\cite{olesinski_ag-si_1989}. The difference can also be observed in the reflectance spectra in Figure~\ref{fig:4}e,f. Together with the change on the pathway to the instability at 600~$^\circ$C, these results confirm the dependence of the thermal stability on the stacking order.    

\section{Conclusion}
If hyperbolic metamaterials are to become widely used in fields such as thermophotovoltaics or radiative cooling, it is crucial to understand their thermal stability. We examine magnetron sputtered \ce{Ag}/\ce{a-Si} layered HMMs as a model system to study the limits of stability at elevated temperatures. Using interfacial roughness as design parameter, we achieved near-perfect absorption above 0.45 homologous temperature in the studied HMMs by coupling to non-radiative modes.
\\~\\
We show that the system undergoes a thermal instability upon annealing in vacuum at 300~$^\circ$C. We investigate the driving forces behind this structural degradation by a combination of FIB nanotomographies and FEM simulations. Our results indicate that the main driving force is the inhomogeneous elastic strain energy caused by a mismatch in thermal expansion between the metallic phase of the metamaterial and the substrate. This contribution has often been disregarded~\cite{cunningham_unraveling_nodate,lewis_stability_2003} or assumed to be homogeneous~\cite{sridhar_multilayer_1997}. Notably, the observed instability is tightly related to the HMM design, as co-sputtered \ce{Ag}/\ce{a-Si}~\cite{galinski_disordered_2019} preserve their structural integrity in the studied temperature range.
\\~\\
We believe that our experiments can open new ways to design metamaterials with enhanced thermal stability. In this context the identification of elastic strain energy as the driving force of thermally-induced degradation in multilayered HMMs can be used to select the appropriate materials for applications requiring near-perfect absorption above that temperature. Besides having the required optical properties, the constituents should be immiscible and have similar CTEs. Other interesting approaches to limit elastic strain driven degradation include geometrical designs limiting the contact area with the surface~\cite{zhao_hyperbolic_2021,riley_near-perfect_2017,wohlwend_chemical_2022} and the use of natural hyperbolic materials~\cite{wang_perfect_2018}.

\section{Experimental Section}
\threesubsection{Synthesis and annealing}\newline
The hyperbolic metamaterials in this work were deposited using magnetron sputtering (\emph{PVD Products Inc}). Silicon (99.99\% \emph{MaTecK GmbH}) was sputtered using radio frequency (RF) sputtering, while silver (99.99\% \emph{MaTecK GmbH}) was deposited using direct current (DC) sputtering. Since DC deposition rates are higher than those for RF~\cite{OhringMilton2002MSoT}, a precise determination of the sputter rates for both materials is needed to fabricate multilayers with the appropriate metal-to-semiconductor ratios. The calibration of the sputter rate was achieved using focussed ion beam cross~sections (FIB-CS) and atomic force microscopy (AFM) experiments. The samples were annealed for 1~h or 12~h in a vacuum at 1 $\times$ 10$^{-9}$ mbar in a \emph{Createc} rapid-thermal annealing (RTA) setup. The heating rate was kept at 5~K/min and no active cooling was used.
\\~\\
\threesubsection{Structural characterization and simulations}\newline
Single and multiple cross-sections of the HMMs were cut, polished, and imaged using a \emph{Zeiss NVISION 40} FIB. The energy-dispersive X-ray (EDX) spectra were acquired using a \emph{EDAX Pegasus} XM 2 System. 
Detailed information on the FIB nanotomography are given in the Supporting Information (Figure~S7). 3D structural mechanics FEM simulations were carried out using Comsol Multiphysics 5.6.
\\~\\
\threesubsection{Optical characterization and calculations}\newline
The refractive index of \ce{Ag} and \ce{a-Si} was measured by Variable Angle Spectroscopic Ellipsometry (VASE) with a \emph{J.A. Woollam} M-2000 system. The optical properties of a-Si are known to be sensitive to the fabrication conditions~ \cite{pierce_electronic_1972,van_dong_electrical_1983}. A comparison between the experimental n and k values of magnetron sputtered \ce{Ag} and \ce{a-Si} with the corresponding literature data is shown in Figure~S1. M-2000 was also used to measure the angular and polarization-dependent reflectance spectra. The reflectance spectra at near-normal incidence were measured using a fibre-coupled reflectometer (OceanOptics).
\\~\\
Optical simulations were performed using Comsol Multiphysics 5.6. The optical phase diagram in Figure~\ref{fig:1}b is based on the effective permittivities calculated using FEM simulations and S-parameter retrieval~\cite{smith_electromagnetic_2005}. Transfer Matrix (TM) calculations were performed in a \emph{Wolfram Mathematica}. We refer to the Supporting Information for a more detailed explanation of the Transfer Matrix Method~\cite{linton2009wave} and how it was modified to account for the effect of roughness~\cite{cozza_optical_2016,szczyrbowski_determination_nodate}. The plasmonic dispersion was calculated based on Ref.~\cite{yeh_electromagnetic_1977}.\end{justify}
\textbf{Supporting Information} \par %Please delete the Suppporting Information statement if it is not applicable. Please supply Supporting Information in another file. Supporting information should not be provided in .tex format
\begin{justify}Supporting Information is available.\end{justify}
% Acknowledgements
\textbf{Acknowledgements} \par %delete if not applicable))
\begin{justify}J.L.O.P thanks for financial support from ETH research grant (ETH-47 18-1). The authors acknowledge the infrastructure and
support of FIRST. Numerical simulations were performed on the ETH Euler cluster. \end{justify}
\begin{justify}

\end{justify}
\end{document}